%% file: template.tex
\documentclass[journal]{vgtc}                     

\onlineid{4181}

\vgtccategory{Research}

\vgtcpapertype{application}

\author{%
  \authororcid{Wei Zhang}{0000-0002-8321-4607},
  \authororcid{Wong Kam-Kwai}{0000-0002-2813-1972}, 
  \authororcid{Yitian Chen}{0009-0000-0590-594X},
  \authororcid{Ailing Jia}{0009-0004-3631-4690}, Luwei Wang,\texorpdfstring{\\}{}
  \authororcid{Jian-Wei Zhang}{0000-0001-8358-6278}, 
  \authororcid{Lechao Cheng}{0000-0002-7546-9052}, 
   Huamin Qu, and Wei Chen
}

\authorfooter{
  \item W. Zhang, Y. Chen, A. Jia, L. Wang, J. Zhang, L. Chen, and W. Chen are with the State Key Lab of CAD\&CG, Zhejiang University. W. Chen is also with the Laboratory of Art and Archaeology Image (Zhejiang University), Ministry of Education, China. E-mail: \{zw\underline{~~}yixian, oscarchen, jiaailing, ppwlwpp, zjw.cs, liygcheng, chenvis\}@zju.edu.cn. Wei Chen is the corresponding author.
  \item W. Kam-Kwai and H. Qu are with HKUST. Email: kkwongar@connect.ust.hk and huamin@cse.ust.hk.
}

\input{content/0_abstract.tex}

\graphicspath{{figs/}{figures/}{pictures/}{images/}{./}} 

\usepackage{tabu}                      
\usepackage{booktabs}                  
\usepackage{lipsum}                    
\usepackage{mwe}                       

\usepackage{mathptmx}                  

\newcommand{\clocation}[1]{\textcolor[RGB]{94, 128, 184}{#1}}
\newcommand{\cperson}[1]{\textcolor[RGB]{88, 138, 138}{#1}}
\newcommand{\ctime}[1]{\textcolor[RGB]{202, 172, 83}{#1}}
\newcommand{\cthing}[1]{\textcolor[RGB]{181, 187, 114}{#1}}

\usepackage{xspace,xpunctuate}

\newcommand{\ie}{\textit{i.e.},\xspace}
\newcommand{\etal}{\xspace\textit{et al.}\xspace}
\newcommand{\eg}{\textit{e.g.},\xspace}

\usepackage[svgnames]{xcolor}
\usepackage{wrapfig}

\newcommand{\rc}[1]{\textcolor{black}{#1}}

\newcommand{\add}[1]{\textcolor{black}{#1}}

\definecolor{Others}{HTML}{AAB0BE}
\definecolor{Political}{HTML}{7891AA}
\definecolor{Academic}{HTML}{AD7982}
\definecolor{Social}{HTML}{8AA79B}
\definecolor{Kinship}{HTML}{CA9087}
\definecolor{Paint}{HTML}{D6BF9E}
\definecolor{Painter}{HTML}{A9906D}
\definecolor{Location}{HTML}{6788BD}
\definecolor{Person}{HTML}{71A6A6}
\definecolor{Time}{HTML}{CAAC53}
\definecolor{Thing}{HTML}{B8C06A}
\definecolor{linkeddata}{HTML}{000000}
\definecolor{biography}{HTML}{000000}
\definecolor{handscrollfeature}{HTML}{000000}
\definecolor{hovera}{HTML}{000000}
\definecolor{hoverb}{HTML}{000000}
\definecolor{hoverc}{HTML}{000000}
\definecolor{hoverd}{HTML}{000000}

\newcommand{\eventtype}[2]{\includegraphics[height=\fontcharht\font`\B]{figs/eventtype/#1.pdf}\textit{\textcolor{#1}{#2}}}

\usepackage{amsmath}
\usepackage{bm}

\mathchardef\dash="2D

\begin{document}



\maketitle

\input{content/1_Introduction.tex}

\input{content/2_Related_Works.tex}

\input{content/3_Background.tex}

\input{content/4_Model}

\input{content/5_Visual}

\input{content/6_Evaluation}

\input{content/7_Discussion}

\input{content/8_Conclusion}

\acknowledgments{%
  The authors wish to thank Peiyi Jiang, Hangchao Gong and Ollie Woodman at Zhejiang University, the editors at CBDB, and Thomas P. Kelly and Yukio Lippit at Harvard University for their valuable feedback and contribution to this project. This work was supported by the Fundamental Research Funds for the Central Universities (226-2022-00235), National Natural Science Foundation Key Project (62132017).%
}


\bibliographystyle{src/abbrv-doi-hyperref}

\bibliography{template}


\appendix 

\end{document}

%% file: content/0_abstract.tex
\UseRawInputEncoding

\abstract{The study of cultural artifact provenance, tracing ownership and preservation, holds significant importance in archaeology and art history. Modern technology has advanced this field, yet challenges persist, including recognizing evidence from diverse sources, integrating sociocultural context, and enhancing interactive automation for comprehensive provenance analysis.
In collaboration with art historians, we examined the handscroll, a traditional Chinese painting form that provides a rich source of historical data and a unique opportunity to explore history through cultural artifacts. 
We present a three-tiered methodology encompassing artifact, contextual, and provenance levels, designed to create a ``Biography'' for handscroll.
Our approach incorporates the application of image processing techniques and language models to extract, validate, and augment elements within handscroll using various cultural heritage databases.
To facilitate efficient analysis of non-contiguous extracted elements, we have developed a distinctive layout. Additionally, we introduce ScrollTimes, a visual analysis system tailored to support the three-tiered analysis of handscroll, allowing art historians to interactively create biographies tailored to their interests.
Validated through case studies and expert interviews, our approach offers a window into history, fostering a holistic understanding of handscroll provenance and historical significance.
}

\title{ScrollTimes: Tracing the Provenance of Paintings as 
  \texorpdfstring{\\}{} a Window into History}

\keywords{Visual analytics, Digital humanities, Painting analysis, Traditional Chinese painting}

\teaser{
  \centering
 \includegraphics[width=\linewidth]{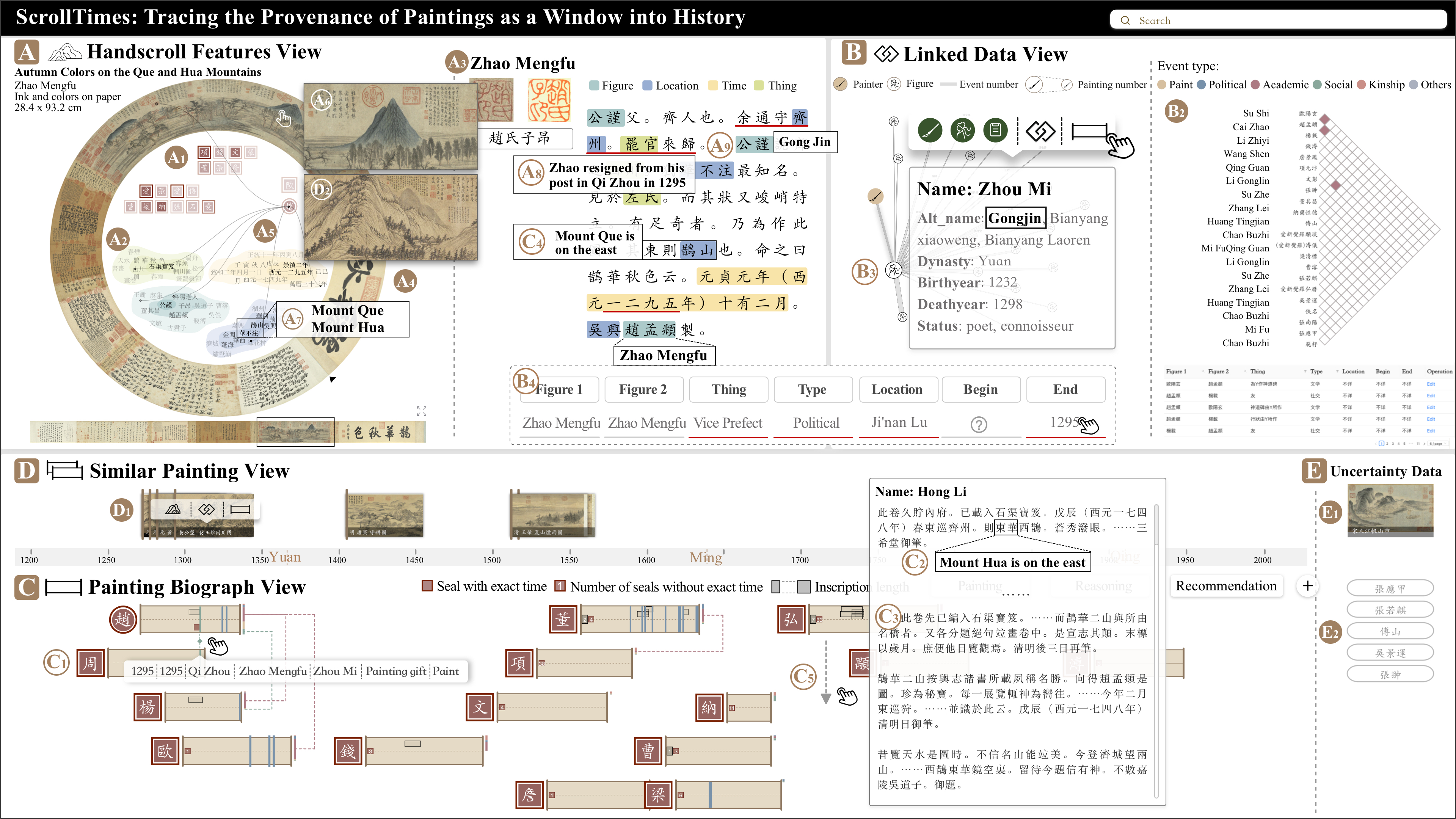}
  \caption{ScrollTimes enables art historians to analyze handscrolls by combining multiple databases and constructing biographies of paintings that fit their research interests. In the case of exploring the renowned handscroll ``Autumn Colors on the Que and Hua Mountains,'' historians first utilized (A) the Handscroll Features View to understand the handscroll elements, such as the core painting, seals, and inscriptions. Then, they further explored the elements of interest in (A3) the Seals and Inscriptions View and validated the information using (B) the Linked Data View. Finally, they constructed a biography of the painting that matched their research interests using (C) the Painting Biography View, (D) the Similar Painting View and (E) the Uncertainty Data View.}
  \label{fig:teaser}
}

%% file: content/1_Introduction.tex
\section{Introduction}

In archaeology and art history, the provenance of cultural artifacts--tracing their ownership and preservation--has been a prominent research focus~\cite{feigenbaum2012provenance}.
Such provenance not only manifests an artifact's authenticity, but also boost its valuation by associating the legacies of previous owners~\cite{newman2011celebrity}.
Moreover, uncovering provenance for cultural artifacts can transcend them from mere static collectibles to vibrant pieces of history, offering new insights and narratives.
With the convenient access to cultural artifact databases~\cite{davis2019old,mohammad2018wikiart,NPM}, digitalized copies of cultural artifacts and their related information have been used to analyze their provenance. 
For instance, ``Mapping painting''~\cite{MP} and ``Hieronymus Bosch''~\cite{HB} visualized the provenance of paintings to engage the wider public in artifacts' preservation efforts.

However, these projects are insufficient in addressing specific challenges faced by professionals in in-depth provenance analysis~\cite{kusnick2021report}.
First, the provenance of cultural artifacts can be documented in other sources and directly on the artifacts. For instance, texture analysis reveals the evolution of style, and inscriptions on the artifacts record changes in ownership. However, recognizing these evidences currently rely on domain expertise, hindering the efficiency of provenance analysis.
Second, each artifact captures the historical context of its era, influenced by its successive owners' narratives. Current research often overlooks the rich sociocultural background against which these cultural artifacts carry during their transmission. 
Consequently, this calls for an integrative approach to connect multiple databases to uncover the intertwined stories of creators, owners, and the historical events they were involved.
Lastly, there is a pressing need to develop sophisticated interactive systems that emphasize automation. These systems should enable researchers to efficiently construct in-depth provenance narratives. Currently, the visual storytelling tools for ``biographies of things''~\cite{kusnick2021report} are limited and often demand excessive manual input.

\begin{figure}[t]
\centering
\includegraphics[width=\linewidth]{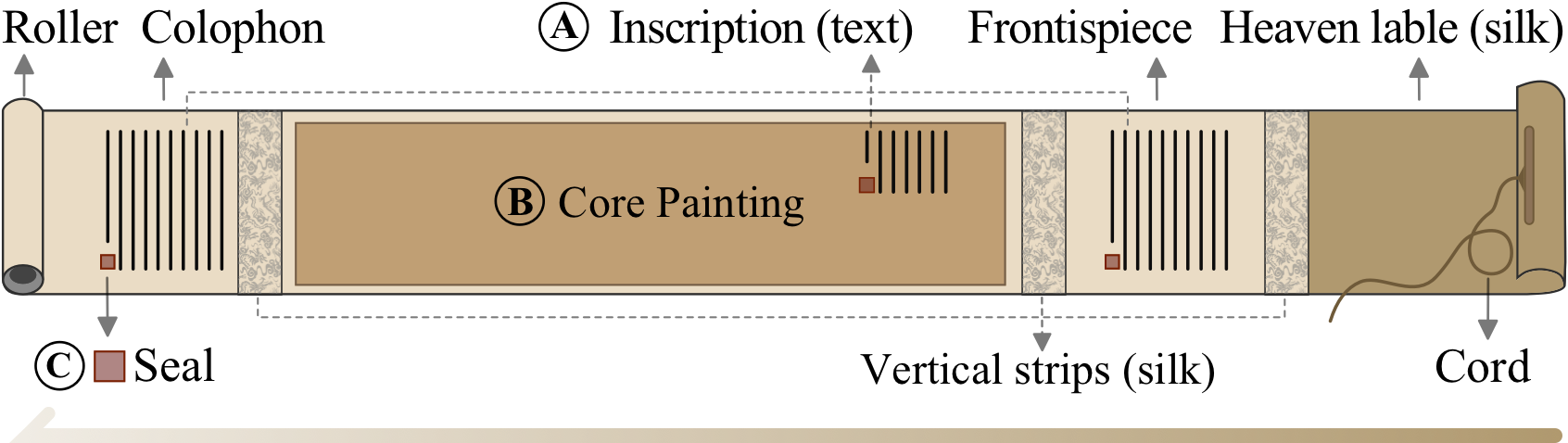}
\caption{Basic handscroll format. It is a long piece of paper or silk that can be rolled up or displayed, and viewers can see different parts of the painting when it is opened from right to left.}
\label{fig:handscroll}
\vspace{-6pt}
\end{figure}

In this study, we explore the provenance analysis on ``handscroll''~\cite{lengyel2000three}, a distinguished format in traditional Chinese painting.
We aim to reveal the historical context embedded in these artworks by combining multiple data sources. 
The handscroll is selected because of its unique creation process, which provides rich historical information over a long period of time. 
The \textit{core painting} (\cref{fig:handscroll}B, usually landscapes or narrative stories) is typically drawn on a long piece of silk or paper from right to left. 
This wide aspect ratio enables the painter to depict the flow of time and the variations in space upon a single handscroll, which can be considered as a scrollytelling.
Upon the completion of the core painting, the painter often adds \textit{inscriptions} (\cref{fig:handscroll}A, text describing the motivation behind the creation) and \textit{seals} (\cref{fig:handscroll}C, vermillion imprint indicating their signature). 
At this point, a handscroll is ``born.''
However, its development does not cease, as the ``life'' of the handscroll continues.
When future generations acquire and appreciate the handscroll, they frequently contribute additional content, such as mounting a new frontispiece or colophon with their own \textit{inscriptions} (including critiques or historical event records) and \textit{seals} (indicating appreciation).
Therefore, the handscroll not only records history but also becomes a part of it~\cite{2003chinese}.

In collaboration with art historians, we propose creating~\textbf{``Biography''} for handscroll to explore their historical narratives. It encompasses three distinct levels:

\begin{itemize}[leftmargin=10pt,noitemsep,topsep=0pt]
    \item{At the \textbf{artifact level}, it describes the characteristics and appearance of cultural artifacts. To improve the efficiency of summarizing the characteristics of the handscroll, we automatically extract the elements embedded within the handscroll. Considering the unique appearance of handscroll, we implement seam-carving~\cite{2007seam} method to optimize layouts facilitating art historians in the focused analysis of non-adjacent extracted elements.}

    \item{At the \textbf{contextual level}, it constructs a social network related to cultural artifacts. We integrate multiple databases to verify extracted elements and supplement information, assisting art historians in uncovering the relationships among \rc{historical figures} and events related to the handscroll.}

    \item{At the \textbf{provenance level}, it chronicles the life experiences of cultural artifacts. We combine \textit{Artifact} and \textit{Contextual} information through temporal visualization. Additionally, we employ a hierarchical algorithm to rank the significance of \rc{historical figures}, allowing art historians to intuitively observe the ups and downs of handscroll's life experiences and compare the features of handscroll provenance across different time periods.}
\end{itemize}

Based on three levels, we propose \textit{ScrollTimes}, an interactive visual analytic system that explores the provenance of cultural artefacts. It automatically generates handscroll biographies and assists art historians in customizing them. We evaluated the system's effectiveness and usability through two case studies and expert interviews. Finally, we have created 1380 handscroll biographies and open-sourced the system at https://scrolltimes.cn.

%% file: content/2_Related_Works.tex
\section{Related Works}

Our research is closely aligned with the topics about the visualization of linked data for cultural heritage and computational methods for analyzing paintings. Consequently, we will now delve into the body of literature surrounding these fields.

\subsection{Visualizing Linked Data for Cultural Heritage}

With the advent of digital cultural heritage databases, open linked data provides a solid foundation for reconstructing the historical evolution of culture~\cite{2019ch, davis2021linked, li2022computing}. Researchers leverage various computational techniques such as semantic web technology~\cite{damova2011reason, dragoni2016enriching}, multi-temporal and spatial reconstruction~\cite{rodriguez20174d, boutsi2019integrated}, and web-GIS~\cite{nishanbaev2021web} to structure and elegantly display the heterogeneous data. These works largely promote resource sharing and comprehensive information exploration. 

Researchers have increasingly integrated multiple cultural heritage databases to discover historical insight~\cite{2023zhangcohort}, enhance access to cultural information~\cite{rodriguez2019diachronic}, and manage multiple small collections~\cite{dragoni2017knowledge}. 
Meanwhile, visualization methods for digital cultural heritage collections have been gaining popularity~\cite{2019ch}, and new applications (\eg Casual Learn~\cite{ruiz2023casual}) and tools (\eg CHEST~\cite{garcia2022chest} and InTaVia~\cite{2022InTaVia}) have been introduced. 
Visual analytics approach has the advantage of contextualizing multi-modal historical knowledge while emphasizing the importance of a user-centered design process.
A variety of heritage items has been explored in depth, including ancient books~\cite{guo2023liberroad}, paintings~\cite{tang2023pcolorizor}, and cultural artifacts~\cite{ye2022puzzlefixer, li2024pmvis}.
In this paper, we focus on paintings in the form of handscroll, which has high cultural values and multiple references in different sources (\eg poems and documentation).

Nevertheless, the credibility of the visual analysis results for cultural heritage remains a challenge due to the intricacy and diversity of cultural heritage data, which often engenders ambiguity and misunderstanding~\cite{zhang2023tcp}. Additionally, current platforms lack better integration of prior domain expert knowledge, which plays a crucial role in professional usages. 
\rc{We improve over existing works in terms of providing art historians with validated entity information from multiple data sources and integrating this information to enable them to make informed inferences based on their prior knowledge.}

\subsection{Computational Approaches for Analyzing Painting}
Recently, a surge of interest has emerged in constructing painting databases from various angles, including the painter~\cite{khan2014painting}, theme~\cite{liong2020automatic,hung2018study, zhang2022visual}, styles~\cite{jiang2019dct}, and even AI-generated paintings~\cite{feng2023promptmagician}. Some support cross-modal retrieval by combining text and image features~\cite{2020tcpdb,6986272}. 
These databases have greatly supported the interpretation~\cite{feng2022ipoet,fan2019evaluation}, creation~\cite{2021wang,2020zfq}, and exhibition~\cite{2013it,2020vr} of traditional Chinese painting~\cite{zhang2023tcp}. However, they tend to focus more on the aesthetic value of the painting itself. MTFFNet~\cite{jiang2021mtffnet} has been proposed to analyze the painters' painting style. Fan\etal~\cite{fan2019evaluation} focus on discovering painting patterns. Tang\etal~\cite{8113507} reproduce the painting process for better painting education. The historical value of paintings remains underexplored.

In fact, behind these paintings is huge historical information on politics~\cite{burke2001eyewitnessing}, economy~\cite{haskell1993history}, and culture~\cite{chung2016understanding}.
Several studies have presented fundamental historical background information of paintings through diverse methods.
For instance, ``Hieronymus Bosch,''~\cite{HB} presents art history through online multimedia, and ``Mapping Paintings,''~\cite{MP} employs maps to display provenance records. However, they are limited to neither engaging users' prior knowledge nor providing comprehensive information. Besides, David \etal's~\cite{9659547} study on the attribution of paintings also lacks sufficient data verification techniques. These works may not fully meet the research needs of art historians who seek to explore the deeper historical context of paintings.

\rc{This study aims to explore the historical significance of Chinese handscroll paintings. 
Different from existing works, we utilize automated methods to extract elements from the paintings and validate them by cross-referencing external cultural heritage databases. The relevant historical contexts can help reveal their historical values.
We construct biographies for these paintings and employ event sequence visualization techniques~\cite{9497654, 2021lan, SHIRATO202377}, especially in multi-modality~\cite{wong2021taxthemis, wong2023anchorage, he2023videopro}, to provide a comprehensive portrayal of their historical significance.}

%% file: content/3_Background.tex
\section{Background}
\label{sec:background}
This section introduces the domain background and outlines the requirements obtained through in-depth interviews with domain experts.

\subsection{Cooperating with Art Historians}
\label{cooperate}
Over the past year, we have engaged in a close collaboration with six highly experienced art historians. This group comprises four professors (A1-A4), each boasting more than twenty years of research experience, and two Ph.D. students (A5 and A6) with nine years of practical expertise.
A1 and A2 specialize in the study of painting during the Song and Ming Dynasties. A3 and A6 concentrate on stylistic analysis spanning entire generations, while A4 focuses on the analysis of Zen-themed paintings, and A5 specializes in the analysis of paintings with an ``Elegant Gathering'' theme. The collaborative process can be divided into three key stages as follows:
\begin{itemize}[leftmargin=10pt, nolistsep]
    \item \emph{Requirement Summary and Task Identification.} We obtained insights into the traditional process of cultural artifact provenance analysis through expert interviews, refining requirements and clarifying system design tasks~(\cref{task}).
    \item \emph{System Design and Enhancement.} Experts actively participated in system design, and we iteratively adjusted and improved the system based on their suggestions (\cref{model}, \cref{vasystem}).
    \item \emph{Case Studies and Feedback.} Experts conducted case studies aligned with their research interests, providing valuable feedback~(\cref{eva}).
\end{itemize}

\subsection{Requirements and Task Analysis}
\label{task}
Through interviews with art historians, we have extracted the traditional process of create biography for artifacts, which typically involves \rc{\textit{formal analysis}, \textit{contextual analysis}, and \textit{critical thinking}.} These experts have emphasized that this process is both labor-intensive and time-consuming, especially in the first two phases. We have streamlined the requirements and design tasks into three levels as bellow.

\noindent{\bf \rc{R1: At the artifact level, analyze handscroll characteristics efficiently.}}
Experts emphasize that the first step in analyzing handscrolls is to conduct a formal analysis of the rich visual elements they contain. Extracting and summarizing these elements aids in uncovering hidden transmission information within the artifact.

\begin{enumerate}[label=\textbf{T{\arabic*}}, start={1}, nolistsep]
    \item {\bf Extracting handscroll elements efficiently.} 
    Analyzing handscrolls' visual properties relies on element extraction. Traditional methods, as indicated by A1-A6, are time-consuming, especially for handscrolls like ``Autumn Colors on the Que and Hua Mountains,''  with 168 seals and 36 inscriptions. The system needs an automatic method for identifying and extracting elements.

    \item {\bf Visualizing element correlations.} 
    A3 and A4 indicate that seal impressions and inscriptions are often scattered along the handscroll. However, the wide aspect ratio of handscrolls hampers efficient analysis, especially when dealing with non-adjacent elements. The system should optimize the handscroll layout to help experts establish element connections quickly.
\end{enumerate}

\noindent{\bf R2: At the \rc{contextual level}, establish handscroll social network.} 
A1 and A3 emphasize that artifacts reflect and influence societal and cultural dynamics. Exploring the interactions among artifacts, historical figures, and events deepens our understanding of their historical and cultural significance.

\begin{enumerate}[label=\textbf{T{\arabic*}}, start={3}, nolistsep]
    \item {\bf Identifying \rc{historical figures} associated with handscroll.} 
    Various \rc{historical figures} are linked to handscrolls, including painters, sealers, inscribers, \rc{historical figures} mentioned in inscriptions, and those with similar artistic styles. Automated methods are required to identify these \rc{historical figures}. Additionally, A5 highlights the potential for historical figures to have nickname, requiring the integration of multiple databases for data validation.

    \item {\bf Supplementing contextual information.}
    Historical events related to the identified \rc{historical figures} are central in constructing social network, aiding experts in understanding their relationship in historical context. 
    The system needs to integrate related databases for comprehensive historical context.
\end{enumerate}

\noindent{\bf R3: At the \rc{provenance level}, interactively construct biographies.} 
Experts have indicated the need for a convenient platform to construct handscroll biographies and analyze their transmission characteristics. Simultaneously, they also require multiple analytical dimensions to cater to diverse research demands.

\begin{enumerate}[label=\textbf{T{\arabic*}}, start={5}, nolistsep]
    \item {\bf Constructing handscroll biographies automatically.} 
    Since handscrolls often survive for centuries, art historians need to study their cultural significance at different historical moments. The system needs to provide an automated method to link data between the artifact and contextual level in a chronological manner to construct biographies.
    
    \item {\bf Facilitating interactive exploration based on research interests.} 
    Art historians use varying criteria for creating biographies. For example, A3 focuses on stylistic transmission, while A5 emphasizes thematic similarity.
    They are collectively concerned with the presence of uncertain information within paintings, such as the appearance of \rc{historical figures} in artwork that are absent from historical records.
    The system should enable interactive customization with different perspectives.
\end{enumerate}

%% file: content/4_Model.tex
\section{Models}
\label{model}

In this section, we initially describe the databases we utilized. Then we have developed automated methods for handscroll elements processing and context information processing~(\cref{task}).

\subsection{Data Description}
\label{data_desc}
This study collects 1380 handscroll data from the open dataset platform~\cite{NPM}. Each handscroll contains the following data elements:

\begin{itemize}[itemsep=2pt,topsep=0pt,parsep=0pt,leftmargin=*]
    \item{\emph{Core painting (\cref{fig:handscroll}B):} main content of the handscroll, often depicting a narrative.}
    \item{\emph{Inscriptions (\cref{fig:handscroll}A):} texts on the handscroll, containing painters' records, historical events, and critiques of the painting.}
    \item{\emph{Seals (\cref{fig:handscroll}C):} red seal stamps imprinted on the handscroll by the owner or appreciator, indicating signature or authentication.}
\end{itemize}

\begin{figure}[ht]
\centering
\includegraphics[width=\linewidth]{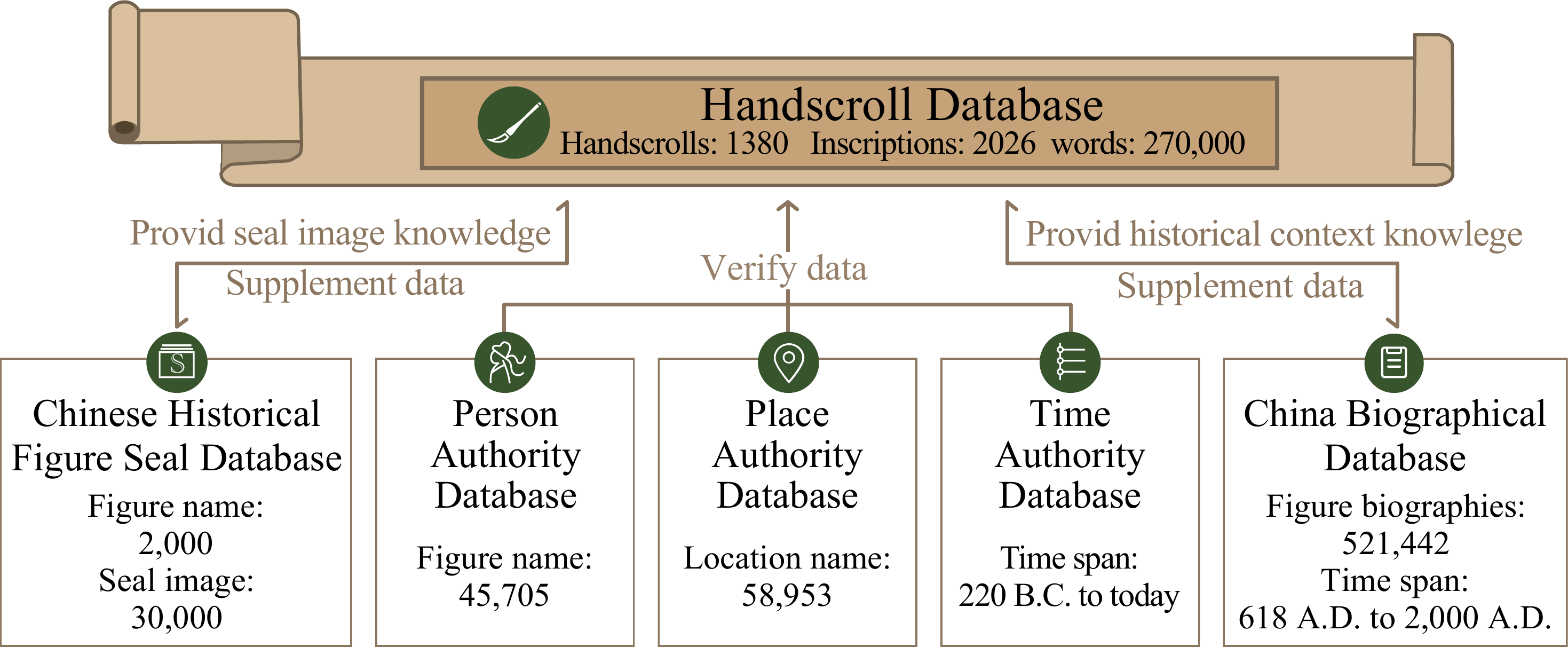}
\caption{The handscroll database is validated and supplemented through the integration of data from multiple data sources.}
\label{fig:database}
\end{figure}

However, these data elements are usually ambiguous and incomplete, which hinders art historians' deep analysis.
To build the handscroll database and perform cross-validation, we selectively collect data from five databases containing information on Chinese historical figures, time periods, locations, and events. These databases (\cref{fig:database}) are chosen based on their relevance to the study of Chinese painting history and their reputation within the academic community. The selected databases include CBDB, CHFSD (Chinese Historical Figure Seal Data)~\cite{CHFSD}, PerAD (Person Authority Database)~\cite{PerAD}, PlaAD (Place Authority Database)~\cite{PlaAD}, and TAD (Time Authority Database)~\cite{TAD}.

\subsection{Handscroll Elements Extraction}
\label{sec:feature_extraction}
At the artifact level, extracting handscroll elements enhances visual properties~\textbf{(R1)}. We use a natural language model for inscriptions and image processing for core paintings and seals~\textbf{(T1, T3)}.

\subsubsection{Inscription Processing}
\label{colophon}
\begin{figure}[htb]
\centering
\includegraphics[width=\linewidth]{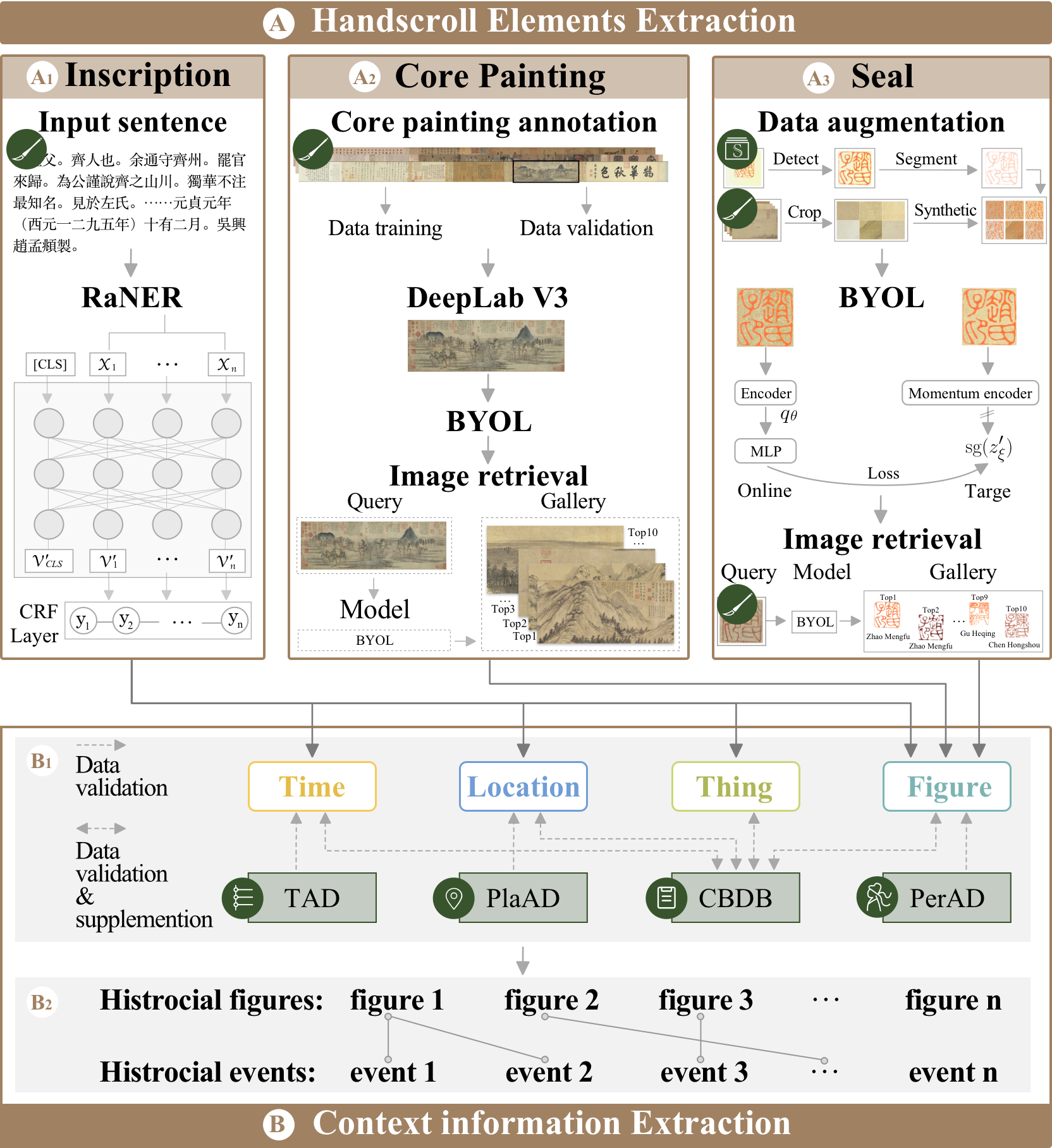}
\caption{Handscroll elements, including (A1) inscriptions, (A2) core paintings, and (A3) seals, undergo processing by various automated models to extract entities. Subsequently, (B1) these entities are validated and supplemented through multiple databases. Finally, (B2) we compile historical figures and events in chronological order.}
\label{fig:workflow}
\end{figure}

The inscriptions used in this study are obtained from the handscroll database. Our objective is to extract valuable information from these inscriptions, including \ctime{Time}, \clocation{Location}, \cthing{Thing} \add{(various objects, such as mountains, birds, and paintings)}, and \cperson{Figure} \textbf{(T1, T3)} (\cref{fig:workflow}A1). 

We evaluate four Chinese Named Entity Recognition (NER) pre-training models, namely BERT-WWM, GuwenBERT, SikuBERT, and RaNER~\cite{RaNER}, to identify entities. We select the RaNER model due to its excellent performance in extracting entities from our dataset.
However, the RaNER model has limitations when it comes to processing lengthy input texts. To overcome this limitation, we implement a sliding window approach~\cite{krystalakos2018sliding} to extract sentences of a manageable length. These sentences are then input into the model for entity recognition. Following this, we employ a simple voting strategy to handle overlapping segments of shorter sentences, ultimately producing the final RaNER classification results.

\subsubsection{Core Painting Processing}
\label{corepainting}

The extraction of the core painting is a crucial aspect of understanding the visual properties of a handscroll~\textbf{(T1)}. Furthermore, since painting places great emphasis on copying and transferring styles~\cite{2018essential}, the inheritance of painting styles can be traced through comparisons of similarities, akin to viewing the ``genealogy'' of a painting (\ie other painters that were influenced by it)~\textbf{(T3)} (\cref{fig:workflow}A2).

In this paper, we employ a deep neural network known as DeepLab V3~\cite{chen2017Rethinking} due to its remarkable performance in semantic segmentation within the field of natural images for extracting the core painting. This method utilizes ResNet~\cite{he2016Deep} as the backbone for feature extraction. To enhance the accuracy of object segmentation, the approach analyzes high-level features extracted from the image and generates multi-scale context features using an atrous spatial pyramid pooling (ASPP)~\cite{he2019road} module. Following the training of DeepLab V3, we utilize it to extract core paintings in the HD dataset, creating a core painting gallery.

To measure the similarity between core paintings, we use a self-supervised learning method called BYOL~\cite{grill2020Bootstrap}. This method involves projecting images into a feature space using an encoder and computing their similarity in that space. During training, two augmented views of each image are fed into an online encoder and a momentum encoder to construct consistency constraints and optimize the encoder. We then retrieve the most similar core paintings from the gallery by comparing the cosine similarities of their query painting features. Then, we obtain core paintings and related painters.

\subsubsection{Seal Processing}
\label{sealprocessing}
Seals play a pivotal role in tracing the provenance of paintings. Our approach to handling seals encompasses seal detection, extraction, and matching. Ultimately, we construct a seal dataset that includes seal images and sealers~\textbf{(T1, T3)}  (\cref{fig:workflow}A3).

For seal detection, we employ a neural network named YOLOX~\cite{ge2021YOLOX} to detect seals in the handscroll. YOLOX integrates the benefits of a series of YOLO models, offering exceptional detection speed and accuracy.
In the experiments, we curate a dataset of Song Dynasty paintings and perform annotations on the seals within the paintings. 
The trained model facilitates the detection of seals from handscrolls and their subsequent isolation for seal matching.

For seal matching, we collect a standard seal gallery including 36302 seal images, along with the corresponding author and textual content~\cite{CHFSD}. By leveraging this gallery, we match the detected seals from handscrolls to identify the sealers and content information.
We train another BYOL model with a ResNet-101 backbone for seal feature extraction.
Notice that seal matching performance may be impacted by the intricate and diverse backgrounds of the handscrolls. Moreover, the gallery samples are limited, with each seal in the gallery having only one image. 
Therefore, e augment the gallery by gathering a set of seal backgrounds from handscrolls and randomly substituting the seal backgrounds with for binary thresholding~\cite{otsu1979threshold}.
After finetuning the BYOL with the new dataset, we project seal images into feature vectors for seal matching. 
We employ Local Sensitive Hashing (LSH)~\cite{indyk1998approximate} to search for the most similar seals from the gallery efficiently.

\subsection{Context Information Extraction}
\label{kg}
At the contextual level, it is necessary to build a social network that involving historical figures and related historical events~\textbf{(R2)}. 
We extracted entities from inscriptions, core painting, and seals~(\cref{sec:feature_extraction}). 
In order to establish a robust social network, it is essential to validate these entities~\textbf{(T3)} through the use of multiple databases. Additionally, to enrich contextual information, we need to integrate data from multiple databases~\textbf{(T4)} (\cref{fig:workflow}D).

\textbf{Figure Validation and Supplementation.}
To validate and enhance the identification of historical figures, we cross-reference the PerAD and CBDB databases and show the data provenance as explanations of automated methods~\cite{feng2023xnli}. Nevertheless, owing to the prevalence of aliases and the use of prefixed names~\cite{2023zhangcohort}, such as official titles or ancient Chinese locations, certain entity recognition results cannot be directly mapped to the names in the database. To tackle this challenge, we employ the following strategies:

\begin{itemize}[leftmargin=10pt, nolistsep]
    \item \emph{Querying PerAD and CBDB directly.} While CBDB provides extensive descriptions of historical figures, PerAD contains valuable data related to religious figures and deities. For example, the painter \cperson{Zhao Mengfu} can be found in both databases, but the named entity \cperson{Jin Mu} is exclusively available in PerAD, where it functions as an alternative name for the \cperson{Queen Mother of the West}~\cite{benard2000goddesses}.

    \item \emph{Querying with sliding-window segments.} For instance, ``Han Lin Qian Pu'' cannot be directly queried. Therefore, by sliding the window in reverse order, it was split into the following words: ``Lin qian Fu,'' ``Han lin Qian,'' ``Qian Fu,'' ``Lin qian,'' ``Han lin.'' After searching in order, it was found that ``Qian Fu'' can be queried, so the record matching result is \cperson{Qian Pu}.

    \item \emph{Determining priority for same name.}
    When encountering duplicate names in the query results, we employ a selection process guided by art historians. This process involves two primary rules:
    First, Era-Based selection. For instance, an inscription from the Yuan Dynasty mentions a figure named \cperson{Gong Jin}, and there are more than twenty historical figures in the database with the alias \cperson{Gong Jin}, we prioritize figures who lived during the Yuan Dynasty, such as \cperson{Zhou Mi} and \cperson{Li Kezhong}.
    Second, Social Relationship criteria. For instance, \cperson{Zhou Mi} has 23 social connections with other confirmed figures mentioned in the painting, whereas \cperson{Li Kezhong} has no such relationships. We identify the figure as \cperson{Zhou Mi} due to the stronger network of associations.
   
    \item  \emph{Allowing interactive figure selection.} While automated methods are beneficial, they may not ensure precise historical figure identification. Hence, art historians can perform fuzzy searches on names and make selections based on the provided list and their expertise. 
\end{itemize}

\textbf{Location Validation.}
Challenges arise due to aliases, prefixes, and suffixes in place names. Similar to the historical figures' process, we cross-reference PlaAD and CBDB databases for matching place entities. For instance, \clocation{Qizhou} is found in both databases, while \clocation{Que mountain} is only available in PlaAD.

\textbf{Time Standardization.}
In ancient China, time are typically indicated by the era name followed by either the year number under that era or the Sexagenary cycle~\cite{nussbaum2005japan}, a traditional time reckoning system. 
Regular expression matching is used to extract the corresponding era name, numeric value, and Sexagenary cycle name, and the era's start and end dates are queried to calculate the exact year. 
For example, if the start year of the ``Yuanzhen'' era is \ctime{1295}, then the time entity \ctime{Yuanzhen second year} corresponds to the year \ctime{1296}. \ctime{Qianlong Wuchen} denotes the year \ctime{1748}, as the ``Qianlong'' era ranges from \ctime{1736} to \ctime{1796}, with \ctime{1748} being the ``Wuchen'' (a Sexagenary cycle name) year.

\textbf{Thing Supplement.}
Due to the various entity types, such as mountains, birds, and paintings, automated methods encounter challenges in verifying and complementing them. Art historians' expertise is essential for augmenting events with these entities. For instance, if a record in CBDB mentions that someone wrote inscriptions for the painting ``Autumn Colors on the Que and Hua Mountains,'' and an art historian includes the entity in the event, it results in a comprehensive event: \cperson{Qian Pu} inscribed an \cthing{inscription} on the painting \cthing{``Autumn Colors on the Que and Hua Mountains''} in \ctime{1446}.

\textbf{NER Performance Analysis.} To assess the impact of post-processing through database matching on result accuracy, we conducted experiments by comparing the entities obtained after database matching as corrected entities with those identified by the NER process. We annotated a subset of the dataset for testing and evaluated the results using the F1 score as a performance metric. The post-processed NER results achieved an F1 score of 0.558, which significantly outperformed the F1 score of 0.483 obtained from the basic NER results. This experiment demonstrates that post-processing through database matching effectively enhances result accuracy.

%% file: content/5_Visual.tex
\section{Visual Analytic System}
\label{vasystem}

\begin{figure*}[htp]
\centering
\includegraphics[width=1.0\linewidth]{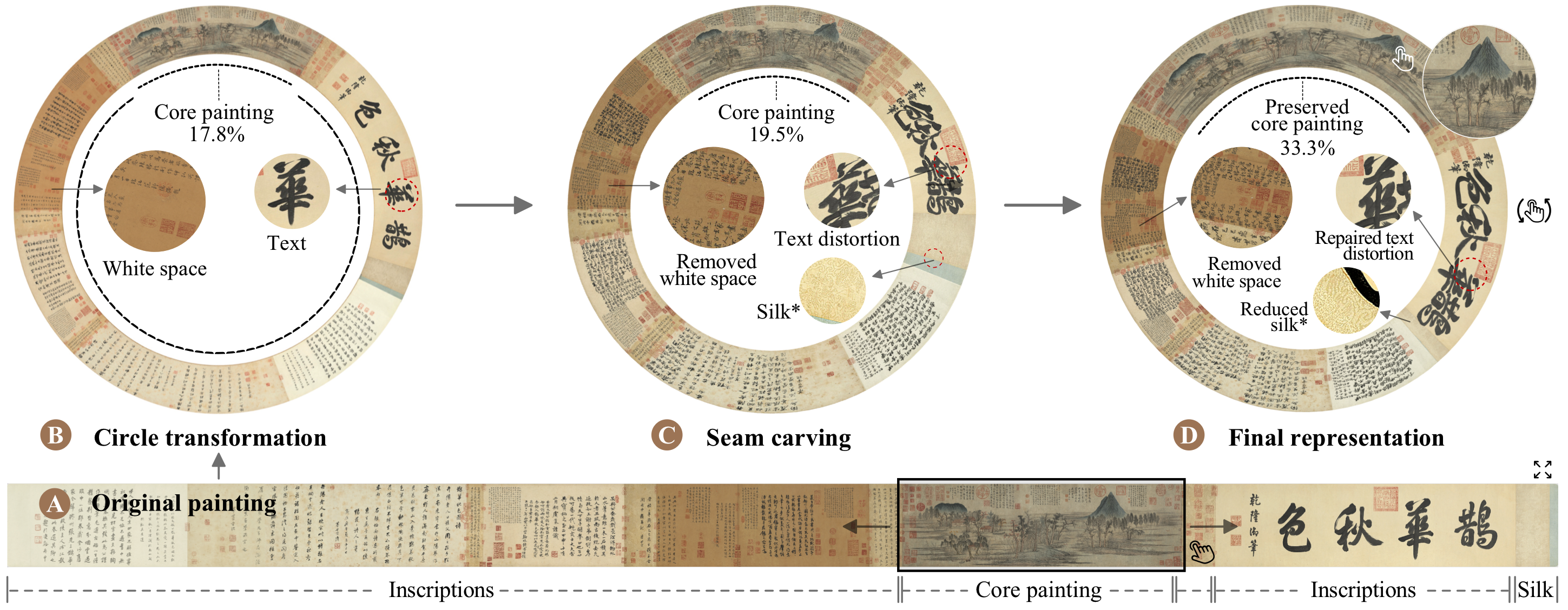}
\caption{(A) The original painting is transformed into (B) a circular ring shape. (C) The seam carving algorithm is used to remove white space. After preserving the core painting, reducing silk, and repairing text distortion, (D) the final representation is displayed. * To better visualize the silk texture, we have applied visual enhancement to the silk in this caption.}
\label{fig:seamcarving}
\vspace{-0.2in}
\end{figure*}

In this section, we propose the visual analysis system, ScrollTimes, which comprises three-tiered components to facilitate the analytical tasks outlined in \cref{task}.

\subsection{Artifact-Level Component}
The Artifact-Level Component (\cref{fig:teaser}A) visualizes the features (\textbf{T1}) and correlations (\textbf{T2}) of handscroll elements, making it accessible for art historians to browse handscroll and gain insights~(\textbf{R1}).

\subsubsection{Summarizing Handscroll Element Features}
\label{sec:wordcloud}
Seals and inscriptions on the handscroll are typically not systematically distributed, often posing reading challenges for art historians due to the ancient Chinese writing style.
To enhance accessibility, we employ a word cloud visualization technique (\cref{fig:teaser}A1, A2) to modernize ancient writing, facilitating user access while accommodating contemporary reading practices~(\textbf{T1}).
Furthermore, \cref{fig:teaser}A3 provides comprehensive details about seals and inscriptions.

\textbf{The Seal Word Cloud View.}
In \cref{sealprocessing}, we extracted the sealers on the handscroll and simplified their visual form into squares (\cref{fig:teaser}A1, \cref{fig:hsdesign}A1). 
The text in the middle of the square represents the sealer's surname, while the transparency of the red background of the text encodes the number of seals they left on this handscroll. 
This allows art historians to quickly understand which sealer left the most seals on the painting scroll. 
The color shade of the outer frame reflects the number of seals they left on the entire handscroll dataset, providing insight into their importance in the history of collecting and appreciation. 
To reflect the changes in dynasty during the transmission of the painting, we cluster the sealers into word cloud visualizaton according to their respective dynasties (\cref{fig:teaser}A1).

\textbf{The Inscription Word Cloud View.}
In \cref{colophon}, we performed entity extraction on the inscriptions. We utilize word cloud visualizations that are systematically categorized by various types of entities, including \eventtype{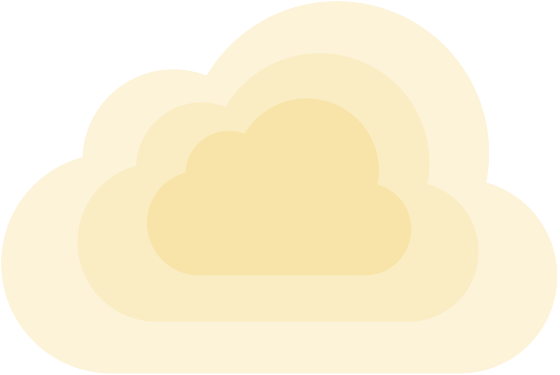}{Time}, \eventtype{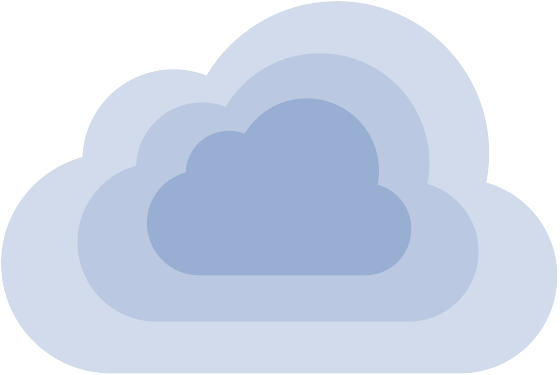}{Location}, \eventtype{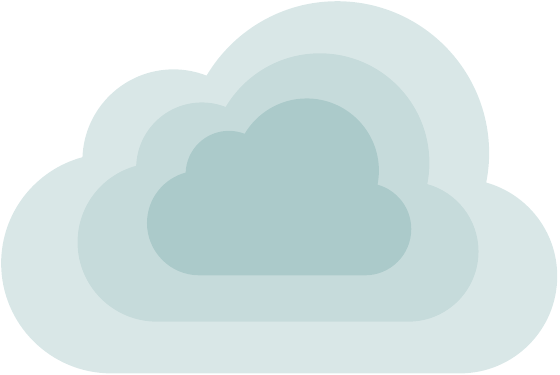}{Figure}, and \eventtype{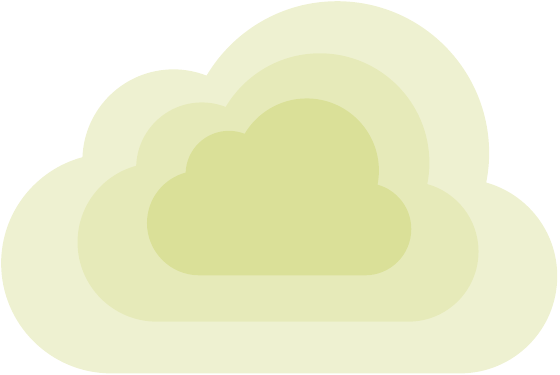}{Thing}. The frequency of words is encoded and represented through discernible variations in font color depth (\cref{fig:teaser}A2).

\textbf{The Seal and Inscription Detail View.} 
This view (\cref{fig:teaser}A3) provides a comparison between the seals on the handscroll and the images in the seal database, along with the original inscriptions that have been identified and labeled by the computer based on entity classification.

\subsubsection{Demonstrating Handscroll Element Correlations}
\label{sec:circle}
Art historians require an accessible method for exploring handscroll and gaining insights from correlations among elements~(\textbf{T2}). We have redesigned the handscroll in a circular format and provide an interactive means of associating elements.

\textbf{The Handscroll Circle View.} This view (\cref{fig:teaser}A4) redesign handscroll considering two principles: 1) displaying the relationships among elements; 2) conserving screen space to highlight the key aspects of the handscroll. Specifically, We demonstrate the design process as follows:

\begin{itemize}[leftmargin=10pt, nolistsep]
    \item \emph{Transforming the handscroll into a circle.}
    The handscroll has wide aspect ratio disparities (\cref{fig:seamcarving}A) that hinder fast retrieval of related contents.
    To maximize the efficient use of system space, we transform them into a circular ring shape (\cref{fig:seamcarving}B).
    The handscroll is divided into two equal-width parts and projected into a 180-degree semicircular ring, which is then flipped and concatenated to form a complete circular ring. 
    To achieve the desired circular effect, each part of the original image is divided into rectangular blocks and proportionally projected into trapezoids based on pixel positions.

    \item \emph{Regulating the ratio of the core painting.}
    In the handscroll, the core painting contains more content and has higher information density than other parts.
    During interviews with art historians, it is determined that the core painting should occupy one-third of the entire length of the handscroll. We resize the handscroll to a fixed length, with one-third being the extracted core painting (\cref{fig:seamcarving}D) and the remaining filled with other parts.
    We adapt the seam carving~\cite{2007seam,2009seam} algorithm for handscroll white space processing (\cref{fig:seamcarving}C). This involves scaling the original handscroll while preserving high-energy content (\ie texts and drawings) in the handscroll and removes low-efficiency information (\ie white space).
    However, seam carving heavily removes the spaces between text characters and distort the appearance (\cref{fig:seamcarving}C). To solve this problem, we divide non-core painting parts into smaller blocks and reduce the compression ratio for blocks with over-deformable texts (\cref{fig:seamcarving}D).

    \item \emph{Reducing silk space.}
    Silk is an inherent part of the handscroll (\cref{fig:handscroll}), but contains relatively less information. However, the complex texture and pattern in the silk material may yield high energy values when calculating gradients, which results in poor squeezing effects on the silk region. Therefore, we introduce the Frequency-tuned (FT) salient region detection algorithm~\cite{2009FT}, which efficiently detects salient objects with clear boundaries and captures the key boundary information of the painting to preserve important content. By linearly combining the FT saliency map and the gradient map, the energy map calculation process is optimized (\cref{fig:seamcarving}D).
    Specifically, the FT algorithm calculates the saliency map based on the Euclidean distance between the pixel colors and the average color in the Lab color space. The input image $I$ is first Gaussian-filtered to remove high-frequency information, resulting in $I'$. With the mean Lab color $\bm{\mu}$ of the original image, we obtain the saliency value in the position $(x, y)$ by $S_{FT}(x,y)=\Vert \bm{\mu}-\bm{c}(x,y)\Vert_2^2,$ where $\bm{c}(x,y)$ is the Lab color of the filtered image $I'$.
    We normalize the saliency maps $\bm{S}_{FT}$ and $\bm{S}_{grad}$ obtained from the FT algorithm and the gradient operator, respectively, and then linearly fuse them to generate a normalized energy map as 
    $\bm{S}=\alpha N(\bm{S}_{grad})+  \beta N(\bm{S}_{FT}),$
    where $N(\cdot)$ is a $\max\dash\min$ normalization function, and $\alpha$ and $\beta$ are weighting coefficients. We empirically set $\alpha=0.7$ and $\beta=0.3$.
    
    \item \emph{Interaction.}
    To aid art historians in focusing on areas of interest from a central perspective, we offer two interactions. 
    First, \eventtype{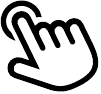}{ }the original handscroll is displayed below the circle view, allowing art historians to select a specific area. The central perspective of the circle view will adjust accordingly (\cref{fig:seamcarving}A).
    Second, \eventtype{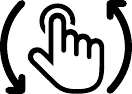}{ } when the mouse hovers over the blank space of the circle view, the ring can be rotated accordingly (\cref{fig:seamcarving}D).
    Additionally, to connect the painting and its elements, we provide interactive lines. 
    Upon clicking on a particular sealer, the solid lines on the view illustrate the distribution of their seals on the handscroll, while the dotted lines indicate the entities mentioned in their inscriptions (\cref{fig:teaser}A5).

\end{itemize}

\textbf{Justification.}  
Beyond its historical significance, the handscroll is a cherished art form. Art historians, in their reverence for these traditional works, question the circular transformation, \add{which does not align with their mental models} and raise concerns about potential damage.
\add{However, when we introduce our handscroll circle design, which efficiently analyzes the characteristics of the handscroll, particularly in connecting scattered elements on the handscroll, art historians greatly appreciate its advantages.} 
Additionally, to cater to art historians' appreciation for the original artwork, we offer two interactive options: hovering over the handscroll\eventtype{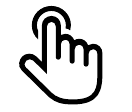}{ } displays the original image of the current section (\cref{fig:teaser}A6), and clicking the zoom-in button \eventtype{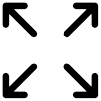}{ } to view details (\cref{fig:seamcarving}A).

\subsection{Contextual-Level Component}
\label{linkdataview}

To provide supplementary information for the handscroll and enable art historians to have a more comprehensive understanding of the painting and conduct relational analysis~\textbf{(R2)}, we extract historical figures and events (\cref{kg}) and present the data through a linked data view (\cref{fig:teaser}B)~\textbf{(T3, T4)}. 

\subsubsection{Exploring Ego Social Networks}
We use the ego network relationship diagram (\cref{fig:teaser}B1), a familiar diagram for art historians, to display the social relationships of individual historical figures~\textbf{(T3, T4)}. To help art historian quickly access painting information, we use a hairbrush icon (\eventtype{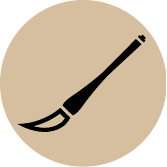}{Painter}) to represent painters, with the icon's size corresponding to the number of associated paintings. 
User can click on a specific \rc{historical figure} mentioned in a seal or inscription (\cref{fig:teaser}A1-A3) to access basic information. Hovering over a figure icon displays their name, birth and death years, nickname, and status, along with associated events in a table below. Clicking on connecting lines between \rc{historical figures} reveals related events in the table, with line thickness indicating relationship strength.

\subsubsection{Exploring Cohort Social Networks}
We present the relationships between \rc{historical figures} through matrix visualization (\cref{fig:teaser}B2)~\textbf{(T3, T4)}.
Cohort analysis is essential for art historians to understand the relationships and interactions between historical figures~\cite{2022zhanguncertainty,2023zhangcohort}. 
Matrix diagram provides a comprehensive and efficient way to visualize these complex relationships and identify patterns. 
This view (\cref{fig:teaser}B2) displays the relationship among the painter, sealer and inscriber of the painting. We also support art historians in adding new \rc{historical figures} to the matrix for further exploration. 
The types of historical events that facilitate further analysis by art historians, we have coded them separately with different colors.

\subsection{Provenance-Level Component}
After recognizing handscroll features and relationships among \rc{historical figures}, historians proceed to understand the handscroll's provenance using the Provenance-Level Component. It offers an automatically generated biography (\cref{fig:teaser}C)~(\textbf{R3, T5}) and supports customization based on their research interests (\cref{fig:teaser}D, E)~(\textbf{R3, T6}).

\subsubsection{Creating Biographies Automatically}
We merge the extracted data from the Artifact-Level and Contextual-Level (\cref{sec:feature_extraction}, \cref{kg}) using temporal visualization (\cref{fig:teaser}C)~(\textbf{R3, T5}). We also apply visual encoding to different data elements to enhance the efficiency of analysis for art historians.

\begin{figure}[htbp]
\centering
\includegraphics[width=\linewidth]{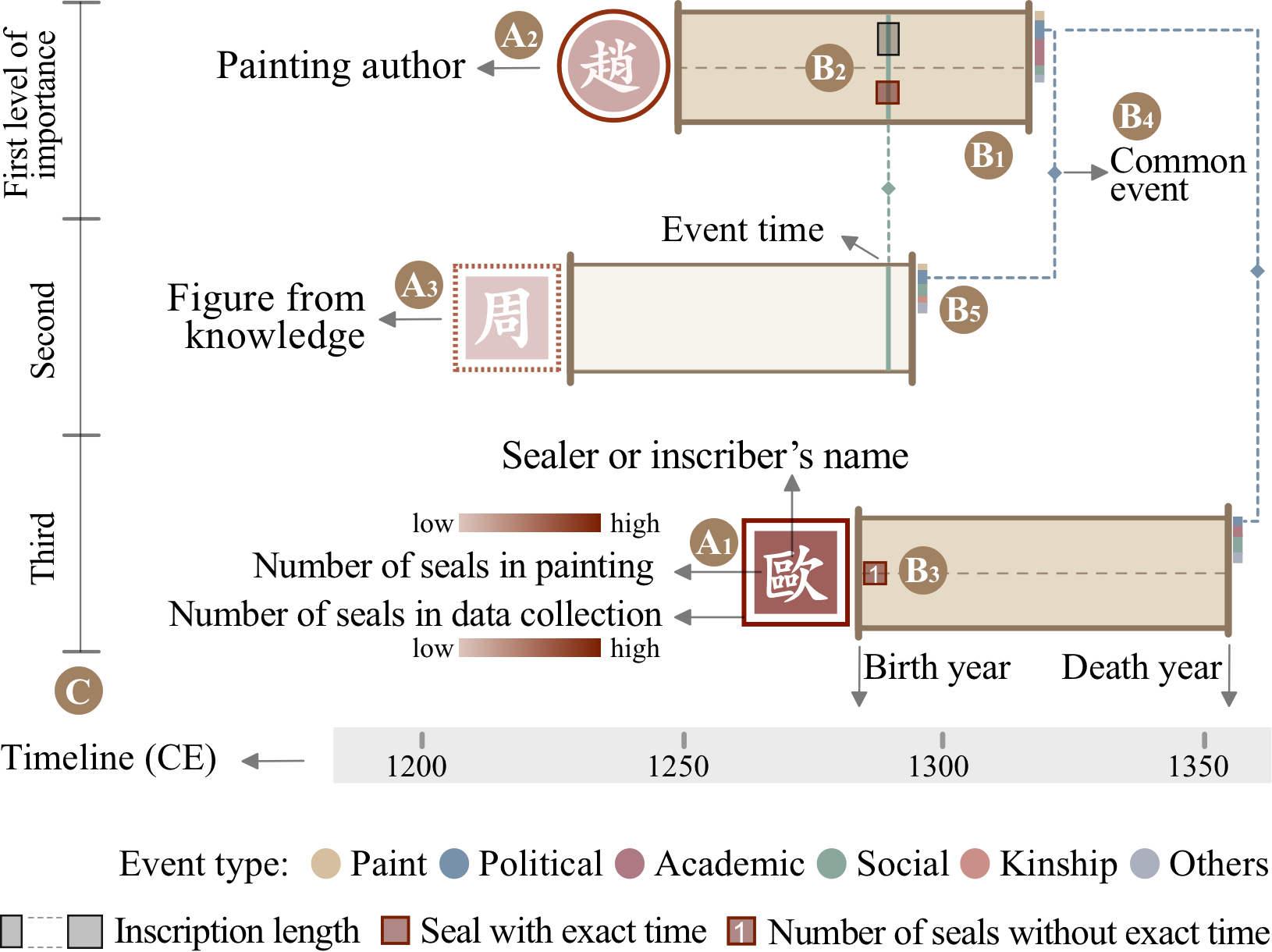}
\caption{Each \rc{historical figure} icon in the Painting Biography View encodes the figure's basic information and its comprehensive level, and related events between figures.}
\label{fig:hsdesign}
\end{figure}

{\bf Mapping \rc{historical figure} type with shape and color.}  
\rc{Historical figures} in the view are categorized into three types. The painting's author is represented by a circular seal (\cref{fig:hsdesign}A2). Collectors and appreciators are denoted by a square seal (\cref{fig:hsdesign}A1) with the same color code as in \cref{fig:teaser}A1. \rc{Historical figures} complemented by art historians have a dashed border (\cref{fig:hsdesign}A3).

{\bf Visualizing the historical figure's lifespan.} 
\rc{For the visual consistency of the entire system, we employ the visual representation of the handscroll to symbolize the historical figure's life span, which we refer to as the ``life-handscroll'' (\cref{fig:hsdesign}B1).}
Above the life-handscroll, we use black blocks to encode inscriptions with exact time, while the size of the blocks encodes the number of words in each inscription. Below the life-handscroll, we use red squares to encode seals with the exact time (\cref{fig:hsdesign}B2). The untimed inscriptions and seals are placed on the left side of the life-handscroll, and the number in the red square represents the number of seals (\cref{fig:hsdesign}B3).
We encode common events as a line with a clickable diamond (\cref{fig:hsdesign}B4) and a stacked bar chart is designed to facilitate the analysis of the distribution and types of events among \rc{historical figures}, which can provide valuable insights into the social and cultural contexts of the \rc{historical figures'} lives (\cref{fig:hsdesign}B5).

{\bf Displaying \rc{historical figure} importance with hierarchy.} 
Under the guidance of art historians, we have developed a comprehensive evaluation metric for determining the importance of figures (\cref{fig:hsdesign}C).
To evaluate a figure's importance score $S$, we consider three key factors:
\begin{itemize}[leftmargin=10pt, nolistsep]
    \item \emph{Painting relevance ($R$).} $R = w_0r_0 + w_1\log r_1 + w_2\log r_2 + w_3\log r_3.$ It contains four aspects, \emph{a)} whether the figure is a pioneer or representative of a painting school, and the importance $r_0$ of the painting scroll, \emph{b)} the number $r_1$ of his paintings, \emph{c)} the number $r_2$ of his paintings being appreciated, and \emph{d)} the number $r_3$ of paintings he appreciated. The coefficients $w_0$, $w_1, w_2$, and $w_3$ are used to balance the four terms. We reduce the weight of Qing Dynasty data in our analysis for balance, due to its excess of paintings.
    \item \emph{Degree of discussion ($D$).} $D = \log d.$ The number of times $d$ the painter is mentioned in ancient literature~\cite{ARS}.
    \item \emph{Figure identity ($I$).} $I = \sum_{i\in\mathcal{I}}s_i.$ where $\mathcal{I}\subseteq$ \{painting collector, literati, official\} is the identity set. The corresponding $s_i$ for painting collectors and literati are both assigned a value of 10, while the $s_i$ for officials ranges from 0 to 20, depending on their official positions. 
\end{itemize}
The complete score of the figure is finally defined by $S = \lambda_1R + \lambda_2D + \lambda_3I,$ where coefficients $\lambda_1,\lambda_2$, and $\lambda_3$ determine the weights of factors. All of the coefficients are set as 1 by default and can be adjusted by the experts according to their preference.
\add{When figure icons overlap, we provide a hover interaction that highlights and pops them out, reducing visual clutter.}

\subsection{Creating Biographies Interactively}
We offer various interaction methods for art historians to create biographies with research interests (\cref{fig:teaser}D, E)~(\textbf{R3, T6}).

\subsubsection{Similar Painting View}

Art historians hope to discover potential painting apprenticeship relationships through the similarity of the painting styles (\cref{fig:teaser}D). The approach outlined in \cref{corepainting} was utilized to obtain data on paintings that exhibit similarities to the original painting.  Meanwhile, some paintings are similar to each other in terms of theme, so we use Term Frequency-Inverse Document Frequency (TF-IDF)~\cite{turney2000learning} model to obtain similar painting data. In order to facilitate art historians in comparing paintings, we set the view position to be close to the original painting and sort the paintings according to the author's birth year.
When an art historian is interested in a similar painting, they can click on the \eventtype{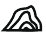}{Handscroll Feature View} button to start a new exploration of the painting. Additionally, if they are interested in the author of a similar painting, they can click on either the \eventtype{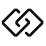}{Linked Data View} or \eventtype{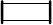}{Painting Biography View} button to explore that painter.

\subsubsection{Uncertainty Data View}
During our interview with art historians, they all expressed interest in the uncertainty data in paintings. Specifically, when processing similar painting data, identifying artworks without artist names and dates (\cref{fig:teaser}E1) can inspire further artistic style analysis. Furthermore, in the figure extraction process, historical figures rarely found in other databases (\cref{fig:teaser}E2) piqued art historians' curiosity. 
They noted that this uncertainty data holds research potential.

%% file: content/6_Evaluation.tex
\section{Evaluation}
\label{eva}
\rc{We invited six collaborating art historians (\cref{cooperate}) and two additional art historians, A7 and A8, to assess the system. These two experts were not part of the system's development. A7 is a Ph.D. student, and A8 is a professor with extensive art history experience.}

\rc{We gave a tutorial on how to explore the ScrollTimes with example-based introductions~\cite{yang2023examples}. (1) Introduction to our motivation and system design (25 minutes).} \add{(2) Free exploration of ScrollTimes using Google Chrome on a PC with a screen resolution of $1920 \times 1080$ (45 minutes). We encouraged participants to use the think-aloud protocol to document their experiences.} \rc{(3) One-on-one interviews to collect their feedback (20 minutes).}
We present two case studies and expert feedback to evaluate the usefulness and usability of ScrollTimes.

\begin{figure*}[ht]
\centering
\includegraphics[width=1.0\linewidth]{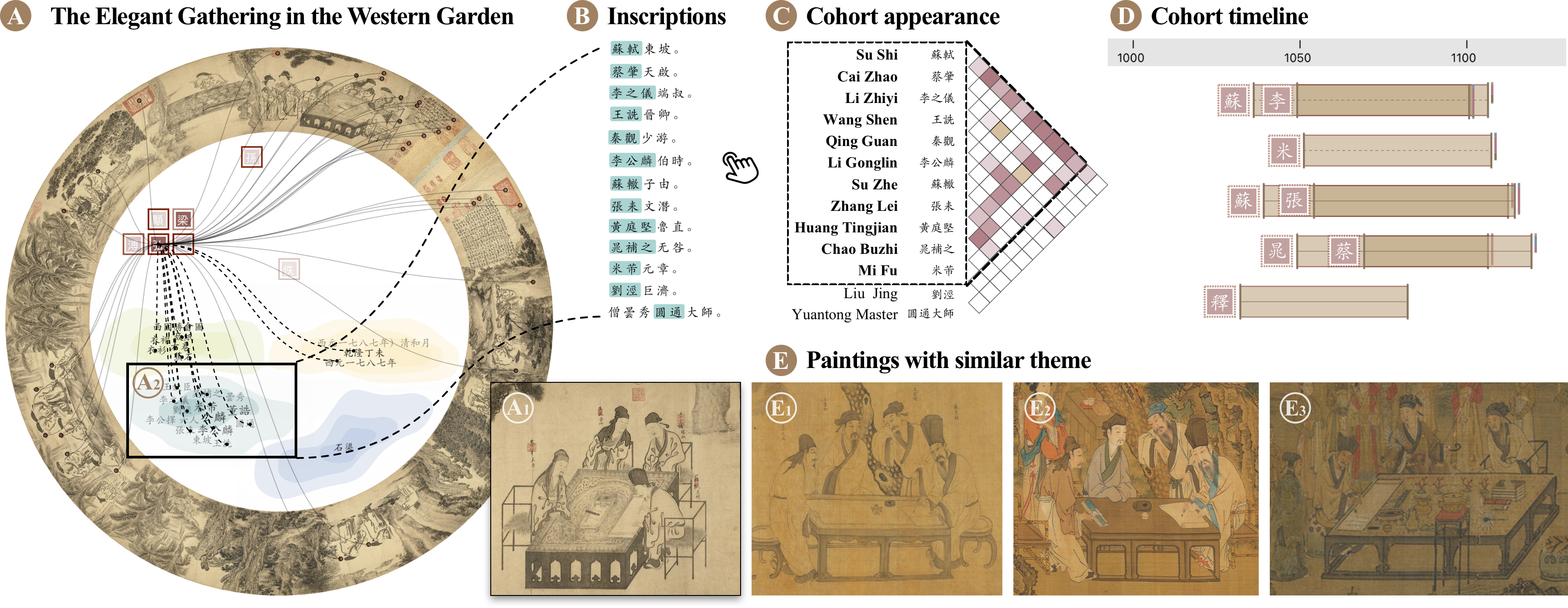}
\caption{Exploration Process. (A) The scene and the figures in the painting ``The Elegant Gathering in the Western Garden'' have been explored. A1 recognizes the main figures' information through (A2) the word cloud and the (B) inscriptions , and (C) the relationship network between the figures indicate that they have formed a close cohort. (D) A1 then explores all figures' life-handscroll to study this idealized gathering. (E) Through the Similar Painting View, A1 finds that ``the Elegant Gathering'' becomes an influential theme in future generations.}
\label{fig:case2}
\vspace{-0.2in}
\end{figure*}

\subsection{Case 1: Autumn Colors on the Que and Hua Mountains}
A1 has a particular fondness for the paintings of Zhao Mengfu, an outstanding painter in the Yuan Dynasty. A1 chooses ``Autumn Colors on the Que and Hua Mountains,'' one of the most extraordinary paintings by Zhao, to dig deep into the painting's biography.

{\bf What does the painting depict?}  
By rotating the circularly-projected handscroll, A1 put the core painting to the top-center of vision (\cref{fig:teaser}A). Consequently, there are mountains with distinct shapes (\cref{fig:teaser}A6). According to the hint through the word cloud below (\cref{fig:teaser}A2), it is revealed that the mountain on the left is \clocation{\textbf{Que} Mountain}, while the mountain on the right side represents \clocation{\textbf{Hua}fuzhu Mountain} (\cref{fig:teaser}A7)~(\textbf{T1, T2}). It is exactly where the origin of the painting's title comes from.

{\bf How was the painting created?}  
To gain in-depth insights into the underlying motivation for its creation, A1 subsequently analyzed the inscription left by the painter \cperson{Zhao Mengfu} (\cref{fig:teaser}A3). It recorded that \cperson{Zhao Mengfu} resigned from his official position in \clocation{Qizhou} (Jinan City, Shandong Province, China) in \ctime{1295} (\cref{fig:teaser}A8). In fact, the linked external dataset CBDB also contains a recorded event corresponding to the description, which, however, lacks the annotations of starting and ending years (\cref{fig:teaser}B4). After combining the two pieces of information, A4 proposed that the missing data could be filled in~(\textbf{T3, T4}). 
It turns out that in-depth analysis based on linked data can even feed back into the complementary of the linked external databases.

\cperson{Zhao Mengfu}'s inscription also showed that this painting was created for a figure named \cperson{Gong Jin} (\cref{fig:teaser}A9), whose frequent appearance in the word cloud also caught A1's interest. However, A1 was not familiar with this figure. By leveraging the Linked Data View, \rc{A1} found that \cperson{Gong Jin} was an alias of \cperson{Zhou Mi}, a celebrated poet of the Southern Song Dynasty known by other aliases \cperson{Bianyang Laoren} (\cref{fig:teaser}B3) (\textbf{T3, T4}). These terms also emerged frequently in the word cloud, which is now known as other references to \cperson{Zhou Mi}. Further inquiry into the historical context of the painting revealed that \cperson{Zhou Mi} and \cperson{Zhao Mengfu} were indeed close friends (\cref{fig:teaser}B3).

{\bf Who collected and appreciated this painting?}
After creation, a famous painting will be widely collected and appreciated by future generations.
\rc{A1} utilized the Biography View to trace the painting's biography from ancient times to the present (\cref{fig:teaser}C) (\textbf{T5}). This view displays the involved historical figures who left seals and inscriptions on the handscroll. As A1 noticed, \cperson{Zhou Mi} was absent since he did not leave any seal or inscription on the painting. However, as previously mentioned, the painting was created for \cperson{Zhou Mi}, which means \cperson{Zhou Mi} still plays a vital role in the painting's biography. A1 thus added \cperson{Zhou Mi} to the Biography View manually (\cref{fig:teaser}C1).

The Biography View implies that this painting has been passed down through the Yuan, Ming, and Qing dynasties and has been collected or appreciated by fifteen historical figures.
Accordingly, a few connections of common events exist among those figures, where the \eventtype{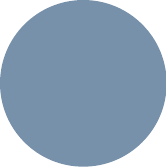}{Political} and \eventtype{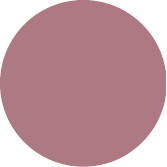}{Academic} relations are most representative.
It is worth noting that three Chinese emperors also left seals on the painting, with Emperor \cperson{Qianlong} leaving up to 33 seals and 9 inscriptions. A1 found something interesting: Emperor \cperson{Qianlong} pointed out a mistake in the painter's inscription. \cperson{Zhao Mengfu} incorrectly stated the position between Que Mountain and Hua Mountain (\cref{fig:teaser}C2, C4). 
However, after examining the content of the inscriptions left by Emperor \cperson{Qianlong} in detail, A1 found that they all revolved around \cperson{Qianlong}'s experience of visiting \clocation{Qizhou} and viewing the \clocation{Que and Hua Mountains} (\cref{fig:teaser}C3). His repeated inscriptions on the same event did not provide much valuable information. Therefore, \rc{A1} chose to lower his importance (\cref{fig:teaser}C5) (\textbf{T6}).
From the above steps, A1 safely drew the collection and appreciation path of this painting, which serves as the most essential part of the painting biography. Nevertheless, A1 tried to investigate the painting genres to complement the biography.

{\bf Who is influenced by this painting?}
The painting style and skills will be passed on, which is also a crucial part of the painting's biography. Specifically, A1 found that a certain similar painting style occurs in other paintings from \cperson{Huang Gongwang} (\cref{fig:teaser}D2), \cperson{Ni Zan}, and \cperson{Tang Yin}, etc (\textbf{T6}). It inspired A1 to take an in-depth look at the Linked Data View, and A1 confirmed that \cperson{Huang Gongwang} did have a teacher-student relationship with \cperson{Zhao Mengfu}. They were in the same painting genre. 
Therefore, A1 added \cperson{Huang Gongwang} to the biography.

So far, A1 indicated that he had a comprehensive understanding of this painting and had constructed its corresponding biography. As A1 stated, this handscroll, which has endured for over 700 years, has been connected to at least 20 influential figures. This handscroll is a time capsule holding many historical records.

\subsection{Case 2: Conversations span the millenniums}

A5 is an art historian who studies figure paintings, and he has a particular fondness for paintings with the theme of ``Elegant Gathering.'' To learn more about this theme, A5 used ScrollTimes to perform a thorough analysis of the masterpiece painting ``The Elegant Gathering in the Western Garden'' to better comprehend the perspective it portrays and the story surrounding it.

{\bf A close cohort of literati.}
A5 rotated the painting circle and observed this painting, in which the figures waved their writing brushes and painted, recited poems and lyrics, and meditated and received Zen instruction (\cref{fig:case2}A1) (\textbf{T2}). A5 looked at the seals and inscriptions extracted and identified from the painting, which didn't arouse his research interest. However, \cperson{Qianlong} left many inscriptions providing important information, that is, the names of the figures in the painting. A5 viewed the word cloud and the inscriptions to further explore the figures in the painting and the story behind their gathering in the Western Garden. Combined with the word cloud (\cref{fig:case2}A4), A5 recognized that the main figures in the painting include \cperson{Wang Shen}, \cperson{Su Shi}, \cperson{Su Zhe}, \cperson{Huang Tingjian}, etc (\textbf{T1, T2}). The inscriptions show detailed information about the figures in the painting, and A5 learned that the story is that \cperson{Wang Shen} invited a painter to make records of the incident where he and his friends gathered in the Western Garden (\cref{fig:case2}B). 
 
The figures in the painting include well-known \cperson{Su Shi} and \cperson{Huang Tingjian}, but A5 was unfamiliar with others, so he placed these figures in the Linked Data View to explore related information (\textbf{T4, T5}). The Linked Data View revealed that every figure in the painting had affairs with \cperson{Su Shi}. The relationship network between the figures indicated that they had formed a close cohort (\cref{fig:case2}C). Further, through event color classification, it can be seen that most of the events occurring within the cohort were \eventtype{Academic}{Academic} events. Among them, there were also \eventtype{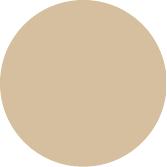}{Paint} interactions between \cperson{Li Gonglin} and \cperson{Li Zhiyi}, \cperson{Wang Shen} and \cperson{Huang Tingjian}. A5 realized that this painting shows a close cohort of literati with \cperson{Su Shi} as the core. This cohort has strong academic resources, promoting literary and artistic innovation since the mid and late Northern Song Dynasty.

{\bf An imaginary gathering.}
Analysis of the close cohort aroused A5's curiosity, so A5 raised a question: Is it possible for so many scholars and celebrities to gather together at the same time? To explore this answer, A5 added figures to the Biography View to examine the possibility of the intersection of this cohort of literati (\cref{fig:case2}D) (\textbf{T6}). The Biography View shows the birth and death dates of all the figures on the painting, all of whom lived in one era (\textbf{T5}). A5 analyzed, ``In terms of time, the figures in the painting were given the opportunity to gather on one day.'' A5 indicated that this form of the gathering was reminiscent of the masterpiece of Western art history, Raphael's ``The School of Athens.'' Raphael did not dwell on the real Plato Academy in history when he painted it; instead, he painted the ideal ``the School of Athens''~\cite{rowland1997intellectual}. So A5 guessed, ``This form of expression may coincide with this painting, although the time allows this gathering, but whether it is a real gathering needs further historical evidence to support.''

{\bf A legacy across the ages.}
In order to explore the impact on other paintings and the potential painting apprenticeship relationships between paintings, A5 used the \textit{Similar Painting View} to observe similar paintings (\cref{fig:case2}E) (\textbf{T6}). Among theme similar paintings, A5 found three paintings with the same theme of ``Elegant Gathering,'' across the Song, Yuan, and Ming dynasties, painted by \cperson{Li Gonglin}, \cperson{Liu Songnian} and \cperson{Tang Yin} (\cref{fig:case2}E). A5 indicated that ``Elegant Gathering'' was a famous theme in the history of painting, so it was natural to find some paintings with similar themes. This showed that later generations did not simply learn and imitate this painting in terms of painting technique and style, but more importantly, they revered and admired the literati collective centered on \cperson{Su Shi}. People's longing for the elegance, spirit, and style of the literati and scholars of the Song Dynasty had led to the repeated imitation and interpretation of ``Elegant Gathering'' by later generations of painters, and the creation is not only a manifestation of their cultural aesthetics but also the ideal world they constructed.

Finally, A5 constructed a painting biography by exploring the stories of the figures and gained a comprehensive understanding of ``The Elegant Gathering in the Western Garden.''

\subsection{Art Historian Reviews}
\label{sec:feedback}

\textbf{Effectiveness of the system.} 
All experts agreed that ScrollTimes significantly benefits painting provenance research, particularly in seal identification and inscription analysis.
A4 highlighted how it streamlines their workflow, transforming a ``manual workshop'' process into a productivity tool, freeing them to focus on more analytical tasks. 
A3 emphasized the time-saving aspect of integrating multiple databases and how ScrollTimes helps pinpoint research entry points. 
Additionally, A1, A2 and A8 expressed enthusiasm for ScrollTimes as a teaching aid in art history classrooms to enhance efficiency and student engagement.

\textbf{Usability of the system.} 
All the experts confirmed the usability of the system. \rc{A5 commented, \textit{``ScrollTimes allows us to visually access handscroll elements features, which was not previously feasible.''} }
A3-5 appreciated the eye-catching circular design of the handscroll (\cref{fig:teaser}A). While the circular format may deviate from the original handscroll's aesthetic, A4 argued that it offers a holistic view of the entire handscroll, and the interaction can preserve the original information.
A8 stated that the Linked Data View (\cref{fig:teaser}B) provided valuable social network information. 
A2 and A3 mentioned that the Cohort View (\cref{fig:teaser}B2) was unfamiliar to them and required some learning effort.
A1 and A4 appreciated the Biography View (\cref{fig:teaser}C), finding it a compelling metaphor representing life through a handscroll. They saw the encoded information as rich and potentially revealing new insights for analysis.
\add{A3 and A8 commented, \textit{``the Uncertainty Data View (\cref{fig:teaser}E) opens up new research perspectives on less common historical figures for us.''}}
A3, A5, and A6 appreciated the Similar Painting View (\cref{fig:teaser}D), believing it could assist them in observing stylistic transmissions. \rc{A7 stated, \textit{``The exploration workflow makes sense to me; it aligns with the traditional workflow and offers numerous useful interactions.''}}

\textbf{Suggestions.}
Art historians provided valuable suggestions based on their research interests. A4 suggested enhancing the comparison of painting biography by supporting the comparison of multiple painting biographies based on the same theme. 
A1-A3 suggested incorporating more ancient book data and displaying the sources of the data.

%% file: content/7_Discussion.tex
\section{Discussion}
\label{sec:discussion}
This section summarizes lessons learned from working with art historians (\cref{cooperate}) and discusses limitations and future work.

\subsection{Design Implications}
\textbf{Open-mindedness.} 
When art historians encounter a new design that surpasses their conventional understanding, it is necessary to showcase the advantage of the new design and
preserve the traditional analytical methods of art historians.
When engaging in interdisciplinary research, it is crucial to recognize that our new proposals may encounter resistance due to violating empirical habits. Therefore, it is important to comprehensively instruct and present the potential advantages of the new methods to collaborative researchers so that they can gain a thorough comprehension, ultimately appreciating the proposed solution.

\textbf{Provenance as narratives.} 
Historians typically specialize in specific periods for an in-depth understanding.
Examining historical events of individuals~\cite{2022zhanguncertainty} or cohorts~\cite{2023zhangcohort} provides a glimpse into short-term historical viewpoints. Through ScrollTimes, we propose to interweave these fragmented viewpoints with cultural artifacts, much like intertwining colorful threads, thereby providing a broader perspective into the evolution of cultural understanding throughout history.
Integrating micro (historical figures) and macro (paintings) perspectives offers new narratives on the long-lasting impact of the short-lived individuals.

\textbf{\rc{Generalizability.}} 
Our workflow for handscrolls leverages multiple models to process and integrate diverse data sources to gain profound insights. 
\add{Moreover, the seam-carving technique can accommodate various handscroll sizes and is intended to inspire more creative designs on the traditional art forms. Its capability in resizing paintings and retaining important details can be transferred to other artefacts to align with specific requirements.}
Other cultural heritage domains, such as the investigation of ancient architectural heritage, can be verified and supported with various multimodal sources (\eg literature and poems).

\subsection{Limitations and Future Work}

\textbf{Aesthetic exploration.}
The ScrollTimes system has primarily focused on exploring the historical value of handscrolls. Still, art historians have pointed out that we should also pay more attention to the aesthetic value of these works and provide analysis protocols on the painting style, brushwork, etc. Although we have included similar paintings in core painting and subject directions, art historians have expressed the need for more detailed similarity calculations based on other artistic styles, such as brushwork and ink, to enhance their analysis and research.

\textbf{Data source expansion.}
Although ScrollTimes has already integrated numerous databases, we are eager to incorporate additional data sources, particularly those related to aesthetic documents, such as ``Empty and Full: The Language of Chinese Painting''~\cite{Cheng1994-CHEEAF} and ``The Chinese on the Art of Painting''~\cite{Osvald2001art} and so on. 
However, these data are recorded as raw texts, lacking a structured database format, making verification and supplementation challenging.

\textbf{Painting biography comparison.}
The ScrollTimes system does not provide an intuitive comparison of painting biographies.
Comparing different biographies can reveal cohort-level cultural heritage activities, thereby enabling analysis and comparisons of heritage styles. 
This will, however, introduce more elements to be compared.
An underexplored direction is to compare paintings in larger displays (\eg immersive devices~\cite{2013it, 2020vr} and combining multiple displays~\cite{tong2023asymcollab}).

%% file: content/8_Conclusion.tex
\section{Conclusion}
This research explores the provenance of handscrolls, representative cultural artifacts with extensive historical content. 
We collaborated with art historians to develop ScrollTimes, which employs a three-tiered methodology (artifact, contextual, and provenance levels) for creating handscroll ``Biography.''
We extract, validate, and enhance elements within handscrolls by integrating data from multiple sources, creating an interactive biography to explore their historical significance.
Two case studies and interviews have demonstrated the system's utility.
In the future, we aim to improve the system for simultaneous handscroll biography comparisons and expand its use to trace the provenance of various cultural artifacts using the three-tiered methodology.